# Title: The green light generation by self-frequency-doubled Yb:YCOB crystal

Authors: Qiannan Fang (1), Haohai Yu (1 and 2), Huaijin Zhang (1 and 2), Jiyang Wang (1)

[1]*State Key Laboratory of Crystal Materials and Institute of Crystal Materials, Shandong University, Jinan 250100, China*
[2]*Corresponding author:* [haohaiyu@sdu.edu.cn](haohaiyu@sdu.edu.cn), [huaijinzhang@sdu.edu.cn](huaijinzhang@sdu.edu.cn)

Taking advantages of the broad emission bands of a $Yb^{3+}$ doped calcium yttrium oxoborate (Yb:YCOB) crystal cut along the optimized direction out of principle planes with the maximum effective nonlinear coefficient, the self-frequency-doubled green light based on the self-frequency-doubling behavior of Yb:YCOB was achieved with a maximum output power of 710 mW at 523 nm.

## 1. Introduction

The green light sources have demonstrated their promising applications, such as laser printing, high-density optical data storage, undersea communications, interferometric measurements, medicine and displays and so on [1]. Self-frequency-doubled (SFD) lasers, combining the active laser medium and the nonlinear frequency conversion medium into a single crystal, have the advantages of more being compact and of lower cost compared with conventional intracavity-frequency-doubled lasers utilising two separate crystals such as Nd:YAG plus KTP. To our knowledge, the monoclinic calcium gadolinium oxoborate ($GdCa_4O(BO_3)_3$, GdCOB) crystal and its family crystal calcium yttrium oxoborate ($YCa_4O(BO_3)_3$, YCOB) are new efficient nonlinear optical and SFD crystals [2-5] with large transparent range, high nonlinear optical coefficient, high damage threshold and high thermal conductivity[6]. Moreover $Yb^{3+}$ ion has a very simple energy level scheme, has no excited-state absorption, and no visible reabsorption loss [7], and the Yb-doped crystals show many advantages with longer radiative lifetimes because there is no concentration quenching effect for a high Yb doping concentration and lower quantum defect that can improve energy conversion from the pump radiation to laser output and reduce greatly the thermal effect in the crystal [8]. Thus Yb:YCOB crystal offers the prospect in the field of SFD materials. Up to now, the highest green output power achieved by an Yb-doped SFD crystal was more than 1 W at 532 nm with Yb:YAl$_3$(BO$_3$)$_4$ crystal [9]. In this letter, the self-frequency-doubled green light was realized with an output power of 710 mW at 523 nm.

2. **Experiment and results**

This Yb:YCOB crystal used for the experiment was grown by the Czochralski method with a doping Yb$^{3+}$ concentration of 20 at.%. It was cut along with their typeⅠphase-matching (PM) direction out of its principal planes because the maximum effective nonlinear optical coefficient is located in their planes[10-12]. The dimensions of the crystal sample for the laser experiment is 4 mm×4 mm×6 mm. The two 4 mm×4 mm faces were polished and the front one was anti-reflection (AR) coated at 976nm and high-reflection (HR, R>99.9%) coated at 1020-1040 nm and 510-530 nm. In order to obtain the maximal absorption, the end face was HR coated at 976 nm and 1020-1040 nm, high-transmission (HT) coated at 510-530 nm.

The experimental equipment was exhibited in Fig.1. The pump source employed in this experiment was a fiber-coupled LD with a center wavelength at 976 nm. The diameter of the fiber is 200 μm with a numerical aperture of 0.22. The pump light was focused into the Yb:YCOB crystal by an imaging unit with a beam compression ratio of 1:1. And the focus length is 30 mm. To transfer the heat from the crystal generated during the laser process, the Yb:YCOB crystal was wrapped with indium foil and mounted in a water-cooled copper block. The temperature of the cooling water was controlled to be 2 ℃. The laser output power was measured by a power meter (1918-R, Newport, Inc.), and the laser spectrum was recorded by an optical spectrum analyzer (HR4000CG-UV-NIR, Ocean Optics Inc.).

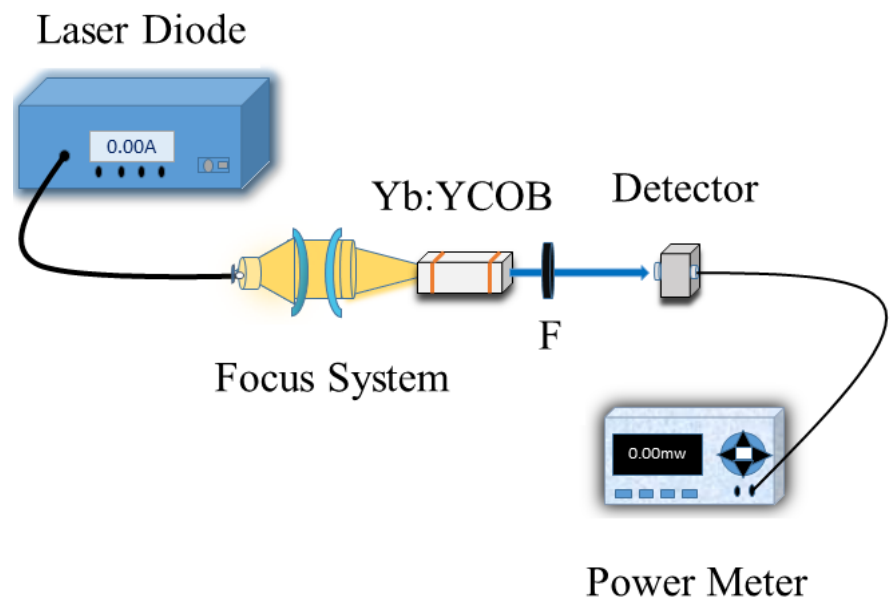

Fig. 1. Experimental configuration of the SFD green light.

With a filter mirror (F) shown in Fig. 1 that is HR coated at 1000–1100 nm and HT coated at 510 -530 nm, the output power performance of SFD green light was measured and detected and was shown in Fig. 2. From this figure, we can see that the threshold was 2.3 W, and the maximum output power was 710 mW under the incident pump power of 10.1 W with an optical conversion efficiency of 8.59%. Upon augmenting the pump power, a crack was found in the Yb:YCOB crystal. With an optical spectrum analyzer, the spectrum of SFD green light was found to be centered at 523 nm, which was presented in Fig. 3.

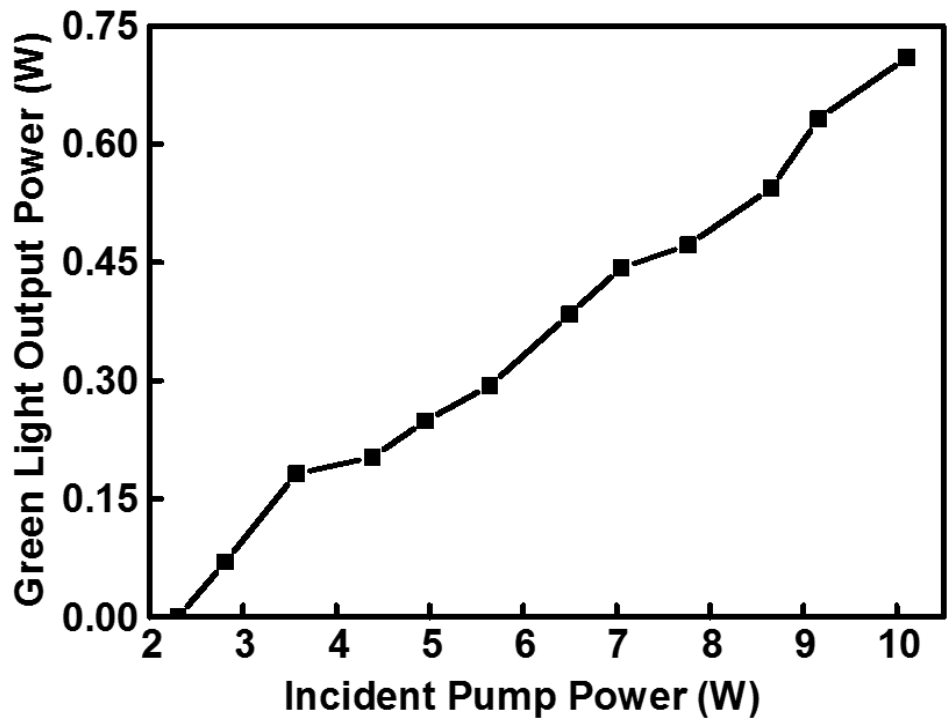

Fig. 2. The SFD green light output powers versus incident pump power.

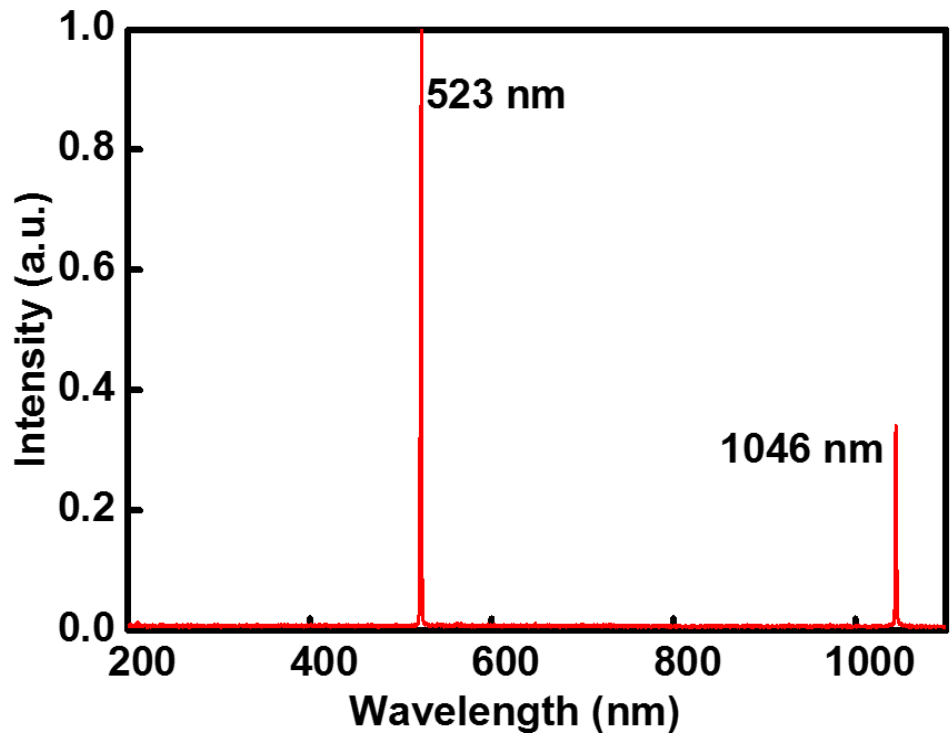

Fig. 3 Output wavelengths of the SFD green light and the fundamental laser.

## 3. Conclusion

By the self-frequency-doubling effect, a maximum green light output power of 710 mW at 523 nm with Yb:YCOB crystal was obtained. This crystal was cut along the phase-matching direction out of principal planes with the maximum efficient nonlinear optical coefficient. This result is the best performance ever reported about the SFD green light with Yb:YCOB crystal. A further improved output power and efficiency would be achieved by optimizing the $Yb^{3+}$ doping concentration and the length of the Yb:YCOB crystal.

**References**

[1] J. Capmany, D. Jaque, J. Garcı_a Solé, and A. A. Kaminskii, "Continuous wave laser radiation at 524 nm from a self-frequency-doubled laser of LaBGeO$_5$:Nd$^{3+}$," Appl. Phys. Lett. **72** (5), 531–533 (1998).

[2] S. Chenais, F. Druon, F. Balembois, G. Lucas-Leclin, P. Georges, A. Brun, M. Zavelani-Rossi, F. Augé, J. P. Chambaret, G. Aka, and D. Vivien, "Multiwatt, tunable, diodepumped CW Yb:GdCOB laser," Appl. Phys. B: Lasers Opt. DOI 10.1007/s003400100477 (2001).

[3] F. Augé, F. Balembois, P. Georges, A. Brun, F. Mougel, G. Aka, A. Kahn-Harari, and D. Vivien, "Efficient and tunable continuous-wave diode-pumped Yb$^{3+}$:CaGdO(BO$_3$)$_3$ laser," Appl. Opt. **38** (6), 976–979 (1999).

[4] F. Mougel, K. Dardenne, G. Aka, A. Kahn-Harari, and D. Vivien, "Ytterbium-doped Ca$_4$GdO(BO$_3$)$_3$ : an efficient infrared laser and self-frequency doubling crystal," J. Opt. Soc. Am. B **16** (1), 164–172 (1999).

[5] D. A. Hammons, M. Richardson, B. H. T. Chai, A. K. Chin, and R. Jollay, "Scaling of longitudinally diode-pumped selffrequency- doubling Nd:YCOB lasers," IEEE J. Quantum Electron. **36** (8), 991–999 (2000).

[6] H. D. Jiang, J. Y. Wang, H. J. Zhang, X. B. Hu, P. Burns, J. A. Piper, "Spectral and luminescent properties of Yb$^{3+}$ ions in YCa$_4$O(BO$_3$)$_3$ crystal," Chem. Phys. Lett. **361** (5-6), 499–503 (2002).

[7] P. Lacovara, H. K. Choi, C. A. Wang, R. L. Aggarwal, and T. Y. Fan,

"Room-temperature diode-pumped Yb:YAG laser," Opt. Lett. **16** (14), 1089-1091 (1991).

[8] H. J. Zhang a, X. L. Meng, L. Zhu, X. S. Liu, R. P. Cheng, Z. H. Yang, S. J. Zhang, L. K. Sun, "Growth and thermal properties of Yb:Ca$_4$YO(BO$_3$)$_3$ crystal," Mater. Lett. **43** (1-2), 15–18 (2000).

[9] P. Dekker, J.M. Dawes, J.A. Piper, Y. Liu, J. Wang, "1.1W CW selffrequency-doubled diode-pumped Yb:YAl$_3$(BO$_3$)$_4$ laser," Opt. Commun.**195** (5-6), 431-436 (2001).

[10] C. T. Chen, Z. S. Shao, J. Jiang, J. Q. Wei, J. Lin, J. Y. Wang, N. Ye, J. H. Lv, B. C. Wu, M. H. Jiang, M. Yoshimura, Y. Mori, and T. Sasaki, "Determination of the nonlinear optical coefficients of YCa$_4$O(BO$_3$)$_3$ crystal," J. Opt. Soc. Am. B **17** (4), 566–571 (2000).

[11] C. Q. Wang, Y. T. Chow, W. A. Gambling, S. J. Zhang, Z. X. Cheng, Z. S. Shao, and H. C. Chen, "Efficient self-frequency doubling of Nd:GdCOB crystal by type-I phase matching out of its principal planes," Opt. Commun. **174** (5-6), 471–474 (2000).

[12] Z. P. Wang, Y. P. Shao, X. G. Xu, J. Y. Wang, Y. G. Liu, J. Q. Wei, and Z. S. Shao, "Determination of the optimum directions for the laser emission, frequency doubling, and self-frequency doubling of Nd:Ca$_4$ReO(BO$_3$)$_3$ (Re = Gd, Y) crystals," Acta Phys. Sin. **51** (9), 2029–2033 (2002).